\documentclass[prd,aps,floatfix,superscriptaddress,twocolumn,10pt,English]{revtex4}

\usepackage{amssymb,amsmath}
\usepackage{graphicx}
\usepackage[percent]{overpic}
\usepackage{overpic}
\usepackage{color}

\newcommand{\Hdim}{H}
\newcommand{\Hadim}{\bar H}

\newcommand{\AvgI}[1]{\left\langle #1\right \rangle_{I}}
\newcommand{\HadimAvg}{\AvgI{\bar H}}
\newcommand{\Pert}[1]{\tilde{#1}}
\newcommand{\funcPrec}{w}

\begin{document}  

\title{
%The dynamics of Schwarzschild EMRI with an external perturber
General-Relativistic Dynamics of an Extreme Mass-Ratio Binary with 
%a Hierarchical Third 
an External
Body
%\MC{Can we really call it an `Inspiral' when we're not including backreaction and so it's just geodesic motion for the internal binary? For a PRL, how about something like `Dynamics of a three-body system in General Relativity'? Would that be too grand/ambitious? } }
}
{
\author{Huan Yang}
\email{hyang@perimeterinstitute.ca}
\affiliation{Department of Physics, Princeton University, Princeton, New Jersey 08544, USA.}
\author{Marc Casals}
\email{mcasals@cbpf.br}
\affiliation{Centro Brasileiro de Pesquisas F\'isicas (CBPF),  Rio de Janeiro, 
CEP 22290-180, 
Brazil.}
\affiliation{School of Mathematics and Statistics and UCD Institute for Discovery, University College Dublin, Belfield, Dublin 4, Ireland.}

\date{\today}

%---------------------------------------------------------------------------------------------------------------------------------------------------------------------------------------------------------------------
%---------------------------------------------------------------------------------------------------------------------------------------------------------------------------------------------------------------------

\begin{abstract} 
We study the dynamics of a hierarchical three-body system in the general-relativistic regime:
an extreme mass-ratio inner binary under the tidal influence of an external body.
The inner binary consists of a central Schwarzschild black hole and a 
 test body moving around it.
We discuss three types of 
tidal effects on the orbit of the test body.
First, the angular momentum of the inner binary precesses  around the angular momentum of the outer binary.
Second, the tidal field drives a ``transient resonance" when 
 the radial and azimuthal frequencies are 
commensurable.
 In contrast with resonances driven by the gravitational self-force, this tidal-driven resonance may boost the orbital angular momentum and eccentricity (a relativistic
 version of the Kozai-Lidov effect).
Finally,  as an orbit-dynamical effect during the non-resonant phase, we calculate the correction to the Innermost Stable Circular (mean) Orbit 
due to the tidal interaction.
Hierarchical three-body systems are potential  sources for future space-based gravitational wave missions
and the tidal effects that we find could contribute significantly to their waveform.
\end{abstract}

\maketitle 

%---------------------------------------------------------------------------------------------------------------------------------------------------------------------------------------------------------------------
%---------------------------------------------------------------------------------------------------------------------------------------------------------------------------------------------------------------------

\section{Introduction}

 The first direct detection of gravitational waves (GWs)~\cite{PhysRevLett.116.061102} by ground-based detectors opens up a window to probe our universe, search for new physics and test the theory of General Relativity with unprecedented means. At the $mHz$ to $Hz$ frequency band, future space-based detectors such as the
 Laser Interferometer Space Antenna  (LISA, see~\cite{lisareview2012,lisa}) will be able to observe signals from 
 extreme mass-ratio inspirals (EMRIs)
 %coalescence 
 of massive black holes and stars/stellar-mass black holes, white dwarf binary mergers, etc. In particular, monitoring the orbital evolution of EMRIs offers a unique opportunity to probe the space-time of a rotating black hole (Kerr), as well as to improve our understanding of the  dynamics of stars in galactic centres~\cite{Peoane2010}.

Because of the separation in mass-scales in EMRI systems, their dynamics can be modelled by black hole perturbation theory.
%for small ratio $\mu$ of the smaller mass to the larger mass, 
%with the least massive object approximated by a point mass.
Within this framework, the small expansion parameter is the ratio $\mu$ of the smaller mass to the larger mass, 
and the least massive object is approximated by a point mass.
 The two-body dynamics can thus be simplified to an effective one-body scenario, where a point mass moves on a geodesic of
 an ``EMRI space-time" whose  metric is the sum of the background metric due to the larger black hole  and the 
  (appropriately regularized) linear~\footnote{Although here we only include the perturbation to first order in 
  $\mu$, 
%  the ratio of the smaller mass to the larger mas,
  that motion is geodesic on the ``background+perturbation" space-time has been shown to second order in~\cite{pound2012second,gralla2012second}; it is expected that it holds to higher orders as well.} gravitational  perturbation generated by the smaller object. 
%  an ``EMRI space-time", defined by the following metric. Using the  ratio $\mu$ of the smaller mass to the larger mass as an expansion parameter, the metric of the EMRI space-time  can be decomposed as the sum of the background metric due to the larger black hole  and the   (appropriately regularized) linear~\footnote{Although in the term ``EMRI space-time" we only include the perturbation to first order in $\mu$,   that motion is geodesic on the ``background+perturbation" space-time has been shown to second order in~\cite{pound2012second,gralla2012second}; it is expected it holds to higher orders as well.} gravitational  perturbation generated by the smaller object. 
That is, the smaller object undergoes geodesic motion on this  ``background+perturbation"~\cite{Detweiler-Whiting-2003,DetweilerPRL2001}. In general, such trajectory is no longer geodesic on the background space-time, and the deviation is due to the metric perturbation induced by the smaller object itself, which give rise to a gravitational self-force~\cite{Poisson2011}. Motivated by future space-based GW missions, understanding EMRI dynamics via the gravitational self-force has been one of the major efforts in gravitational physics in the past couple of decades.

If an EMRI system is not isolated, but is instead  influenced by another massive astrophysical object, e.g., a supermassive black hole, the orbital dynamics and GW radiation are likely modified by the gravitational interaction between the inner binary and the third object. For instance, Yunes et al.~\cite{YunesMiller2011} studied the acceleration of the EMRI system due to the gravitational attraction of the third body
%. Using the effective-one-body formalism~\cite{buonanno1999effective} (involving Post-Newtonian expressions for the flux) in the adiabatic regime for the  inner binary, 
and
they estimated the resulting phase variations in the gravitational waveform. On the other hand, even in the rest frame of the inner binary, the tidal field induced by the third body  changes further the 
EMRI 
%``background+perturbation"
space-time. Such modification was first computed by Poisson~\cite{PoissonPRL2005} for the case of a non-rotating (Schwarzschild) central black hole and later on by Yunes and Gonz\'alez~\cite{YunesGonzalez2006} for the case of a Kerr black hole. 

Understanding the dynamical influence of a tidal field on an EMRI orbit and waveform is the central goal of our work. 
As a first step along this path, we 
consider  an extreme mass-ratio (inner)  binary
%(without including the self-force, but we shall still refer to it as an EMRI)
  within an external tidal field under the following assumptions.
We assume that the tidal field is created by a 
source which is 
slowly-moving around the inner binary (thus constituting an outer binary) and includes only the leading quadrupole moment (since the source is far from the inner binary).
%created  acting on a Schwarzschild EMRI.
As for the inner binary, we assume that the central black hole is a Schwarzschild black hole and we ignore  self-force effects (despite that we shall still refer to it as an EMRI).

Even with a Schwarzschild central black hole and ignoring self-force effects, we discover interesting and new effects due to the tidal field.
Specificall, we   investigate the following   tidal-field effects on the orbit of the smaller particle.
% within this physical system.
%\MC{throughout the paper we use the same term `EMRI system' to indicate an extreme-mass ratio binary when including the SF but also when not including the SF. I think we should use different terms to distinguish between these two cases - Use  `EMRI system' and  `EMRI s-t' without including SF but say that we can trivially add it up whenever we want} 
%which are similar to effects of  the self-force within an EMRI system  (without an external  field). 
%We assume the third body to be static as a first approximation to slow motion.
%We will focus on the three following effects.
First, we  show that the tidal field causes the angular momentum of the  inner binary  to precess around the angular momentum of the outer binary. This precession is caused by the interaction between quadrupole moment of the inner orbit and the tidal field.
%To the best of our knowledge, such precession effect in three-body systems has not yet been observed before.
%within Newtonian or Post-Newtonian theory.
Second, we show that the tidal field leads to transient
 resonances:
 when the ratio of the (evolving) radial and angular orbital frequencies
 is a rational number,
  corrections larger than unity in the orbital phase may occur
and
  the magnitude of the angular momentum may be boosted. 
  %over a timescale $\sim M/\sqrt{\mu}$
 % $\sqrt{\mu}$ times the radiation-reaction scalar, which is $M/\mu$
% although they  lead to
 Equivalent  resonances have been observed within EMRI systems in the absence of a tidal field when including the dissipative piece of the self-force~\cite{flanagan2012transient}.
 However,  in contrast with our case, these self-force-driven resonances
 % although they  lead to corrections larger than unity in the orbital phase, 
% they 
 cannot increase the magnitude
 of the angular momentum (and can only  occur when the central black hole is a Kerr black hole). 
Third and last,
we calculate the shift, due to the tidal field, in the frequency, radius, energy and angular momentum of the Innermost Stable Circular Orbit (ISCO), which, for our system, we define
 in some orbital-average sense.
Equivalent shifts have been found to be caused by the conservative piece of the self-force on particles moving around a Schwarzschild~\cite{BarackSago09} or a Kerr~\cite{isoyama2014gravitational} black hole.
%In this paper, we calculate the tidal-force equivalent of these self-force effects, for the case of a tidal force with only a quadrupole moment created by a static source acting on a Schwarzschild EMRI.
In our case, the frequency shift can be either positive or negative, as opposed to the self-force case which has been found to be positive.

The precession of angular momentum precession may lead to detectable GW phase variations within the precession timescales. 
The resonance effect could have significant observational imprints on the gravitational waveforms as we shall demonstrate later, which we expect to be true for generic Kerr-EMRI.
The ISCO shift affects the peak frequency of the gravitational waveform, but we expect it to be small for the hierarchical triple systems considered here.

It is worth mentioning that in planetary systems, similar three-body dynamics have been extensively studied and many interesting behaviours have been discovered in the Newtonian and Post-Newtonian regimes
% \cite{YunesMiller2011} doesn't fall in the PN case (via EOB) because it's simple Doppler shift, no dynamics of 3-body syst.
 For example, the well-known Kozai-Lidov (KL) mechanism~\cite{kozai1962secular, lidov1962evolution} suggests that the inner binary could trade eccentricity for inclination angle under the influence of the quadrupole tidal field of the third body. In recent years, the KL mechanism has been further extended to include eccentric orbits~\cite{lithwick2011eccentric}, the octupole tidal field~\cite{li2015implications,katz2011long} and Post-Newtonian corrections \cite{WillCQG2014, naoz2013resonant}. As the inner binary transfers from the Newtonian regime to the relativistic regime, the degeneracy between radial and angular orbital frequencies breaks down, which in principle allows for a much richer  phenomenology, as indicated by previous Post-Newtonian studies \cite{naoz2013resonant}. To the best of our knowledge, the present work is the first study of the dynamics of such three-body systems in the fully relativistic regime.
%Most of such systems reside in the Newtonian regime, 
%Specifically, we calculate blabla.

%---------------------------------------------------------------------------------------------------------------------------------------------------------------------------------------------------------------------

\subsection{An order-of-magnitude analysis}\label{sec:order}

Before moving into a detailed analysis in later sections, we first present an order-of-magnitude estimate of the relative strength between the tidal field and the smaller object's self-force.
Such analysis may serve as an indication of the orbital modification generated by the tidal field during the gravitational radiation-reaction timescale.
Throughout the paper we use units with $c=G=1$.

Let us denote the inner binary orbit separation by $r_0$ and its two masses by $M$ and $\mu M$, with $\mu \ll 1$.
The third body is at a distance $d$ and it has a mass $M_*$. 
The system is illustrated in Fig.\ref{fig:system}.

\begin{figure}[h!]
\begin{center}
  \includegraphics[width=4cm]{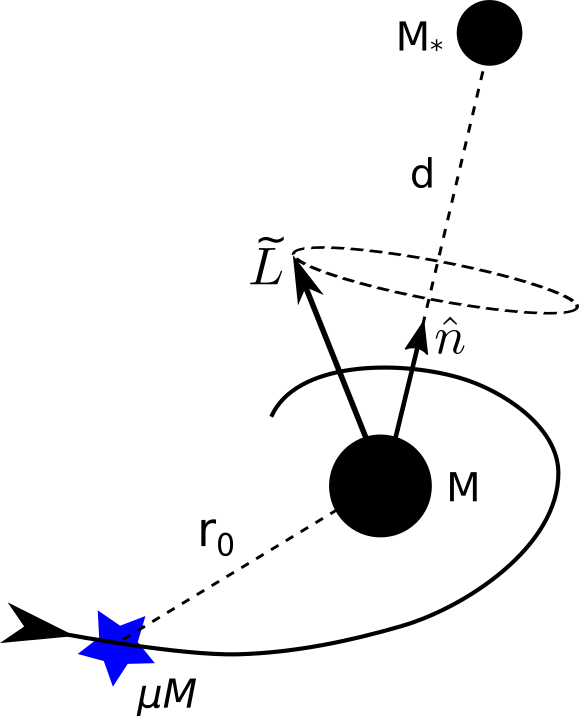}
    \includegraphics[width=4cm]{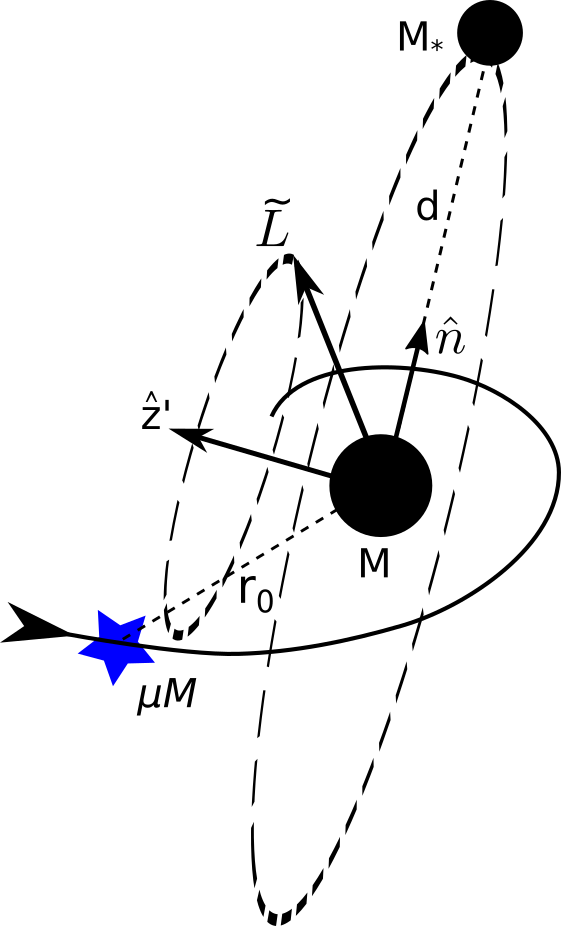}
\end{center}
\caption{
Illustration of our three-body system: an inner binary composed of the larger black hole of mass $M$ and  the 
smaller compact object of mass $\mu M$, and a third body of mass $M_*$.
This third body is distant from  the inner binary and is, generally, orbiting around it.
The angular momentum of the orbit of the smaller object is ${\bf \Pert{L}}$, which is perpendicular to the orbital plane of the inner binary
and, in general, is neither parallel nor perpendicular to the direction between $M$ and $M_*$, indicated
by the unit vector $\hat n$.
We show in Sec.\ref{sec:Precession} that ${\bf \Pert{L}}$ precesses around 
$\hat n$ assuming the tidal field is static within the period of inner binary: left plot.
%an axis $z'$, which is parallel to
%the angular momentum of the orbit of the outer binary.
 In realistic situations, however, we also need to apply the orbital average over the third body. After that averaging we find that $\Pert{L}$ 
 precesses around a vector $\hat{z}'$, which is parallel to the angular momentum of the outer binary: right plot. }
\label{fig:system}
\end{figure} 

The dynamics of the inner binary is influenced by:
(i) the background spacetime of its MBH
of mass $M$;
(ii) the gravitational field of  its smaller body
 of mass $\mu M$ ($\mu\ll 1$;  typically for EMRIs: $\mu\sim 10^{-4}-10^{-8}$), 
causing a gravitational self-force \cite{Poisson2011,pound2012second,gralla2012second,Detweiler-Whiting-2003,DetweilerPRL2001};
 (iii) the tidal force generated by the third body, another MBH
  of mass $M_*$.

The dissipative part of the self-force is the driver of the secular change of the conserved quantities of the orbit
 of the EMRI system
to
$\mathcal{O}(\mu)$.
The self-acceleration is
$a_s \sim \mu  v^9/M \sim  \left ( M/r_0\right )^{9/2}\mu/M$, where
$r_0$ and $v$ are, respectively, the characteristic EMRI orbital separation and speed.
The tidal acceleration is
$a_{\rm tide} \sim M_* r_0/d^3$,
where $d$ is the distance between $M$ and $M_*$.
As we show later, the orbital phase-shift generated by the tidal field during a transient resonance~\cite{flanagan2012transient} is $\sim \mu^{-1/2} a_{\rm tide}/a_s$, whereas LISA's phase resolution of a given event is $\sim 1/{\rm SNR}$ (SNR: signal-to-noise ratio). Therefore, a tidal event is detectable if
%(take $M$ to be the mass of less massive BH of the binary MBH)
\begin{align}\label{eq:distance}
d &\le 0.1 \,{\rm pc}\, \times  \nonumber \\
&\left ( \frac{\mu}{10^{-6}} \right )^{-1/2}\left (\frac{M_*}{ M}\right )^{1/3} \left ( \frac{\rm SNR}{40}\right)^{1/3}\left ( \frac{r_0}{15 M} \right )^{11/6}  \frac{M}{ M_{\rm Sg A^*}} \,,
\end{align} 
  with $M_{\rm Sg A^*} \sim 4 \times 10^6 M_\odot$. We expect the tidal effect is easier to detect around the less massive MBH ($M<M_*$), because its EMRI frequency is closer to the LISA band. Such orbits are also possibly eccentric due to the KL effect, in which case $r_0$ should be viewed as average radius and the peak frequency should be given by the periastron distance. According to \cite{gair2017prospects}, the detection rate of EMRIs by LISA ranges from a few tens to a few thousands per year, if the detection threshold of SNR is considered to be $20$ \footnote{The ``average" SNR of detected events is higher than the detection threshold by definition, but it is not clear what are the exact values from~\cite{gair2017prospects}. Based on Monte-Carlo simulation of binary black holes for LIGO detections in a separate study \cite{yang2017black}, the distribution probability density of SNR roughly follows ${\rm SNR}^{-2}-{\rm SNR}^{-3}$ scaling, and hence the mean SNR of detected events is roughly $2$ times of detection threshold.}. 
  It is believed that tens of percent of Milky-Way-alike galaxies have experienced a MBH
  merger within the past $10$ Gyr~\cite{bell2006merger,lotz2008evolution,YunesMiller2011}. For each merger, the time taken by the MBH binary to migrate to $\sim 1 {\rm pc} $ scale (through dynamical friction) is comparable to the local dynamical time of galaxies ($\sim 10^8 {\rm yr}$)~\cite{kelley2017massive}, but the evolution from
  a distance of $\sim 1 {\rm pc}$  to $\sim 10^{-3} {\rm pc}$ (where GW radiation takes over) is still uncertain (this is known as the final parsec problem \cite{milosavljevic2003long}). Taking the lifetime of MBH binaries to be several Gyr~\cite{kelley2017massive}, it is possible that the decay time starting from a sub-parsec distance (Eq.~\eqref{eq:distance}) is about several hundreds of million years (a few percent of $10$ Gyr). As a result, the optimal detection rate for the tidal effect by LISA  is approximately a few ${\rm yr}^{-1}$.

We organize this paper as follows. In Sec.~\ref{sec:formalism} we review Poisson's~\cite{PoissonPRL2005} approach to calculating the deformation of the Schwarzschild metric due to the presence of an external tidal field. This approach will allow us to subsequently analyze the tidal effect in EMRI dynamics quantitatively.
In Sec.~\ref{sec:secular} we show that the only secular effect by the tidal field outside a resonant phase is the precession of the orbital plane and we compute the precession frequency numerically. We also show in that section that during a resonance phase the rate of change of angular momentum may  be  nonzero. In Sec.~\ref{sec:ISCO} we compute the shifts  in the frequency, radius, energy and angular momentum of an orbital-averaged  Innermost Stable Circular Orbit (ISCO)   due to the tidal field. 
%Through out the paper we adopt the geometric unit that $c=G=1$.  
We conclude with a discussion in Sec.\ref{sec:conclusions}.

%---------------------------------------------------------------------------------------------------------------------------------------------------------------------------------------------------------------------
%---------------------------------------------------------------------------------------------------------------------------------------------------------------------------------------------------------------------

\section{Formalism}\label{sec:formalism}

Our physical setting is that of an
EMRI system, composed of a small compact object
% (modelled as a point particle) 
and a massive black hole, within the influence of an external tidal field.
The small compact object is modelled as a point test  (i.e., the self-force is neglected)  particle and 
it is moving 
%on the background space-time created by the
around a
 massive black hole, which we shall take  to be a Schwarzschild black hole.
The tidal field is  created by a third, remote and slowly-moving (in this paper we take the static limit over the period of inner binary) body; the tidal field is considered
to be a perturbation $h_{\mu\nu}$  of the metric
 $g_{\mu\nu}$   of the massive black hole.
Specifically, in our setting,  $g_{\mu\nu}$   is given by the Schwarzschild metric Eq.\eqref{eq:Schw in Schw coords} below
and the tidal perturbation $h_{\mu\nu}$   will be given later in  Eq.(\ref{eq:eqtide}) combined with Eq.(\ref{eq:electric tensor}).

We may adopt two different but equivalent viewpoints to approach this problem.
We note that these viewpoints apply similarly to the different setting of an EMRI system 
including the  self-force but no external tidal field, in which case $h_{\mu\nu}$ would correspond to the regularized gravitational self-field
($g_{\mu\nu}$ would continue to be the metric of the massive black hole).

In the first viewpoint, the smaller object is moving on an orbit of the background space-time  $g_{\mu\nu}$  and is undergoing an acceleration
due to $h_{\mu\nu}$.
The $4$-acceleration $a^\mu\equiv D u^{\mu}/d\tau\equiv u^{\nu}\nabla_{\nu}u^{\mu}$ is given by~\cite{Poisson2011}
 \begin{align}\label{eq:acc}
a^\mu & = -\frac{1}{2} (g^{\mu\nu}+u^\mu u^\nu)(2h_{\nu \lambda;\rho}-h_{\lambda \rho; \nu}) u^\lambda u^\rho,
\end{align}
where $u^{\mu}=dx^{\mu}/d\tau$ is the $4$-velocity of the particle,
%using the geodesic which instantaneously matches the position and velocity of the accelerated worldline (the osculating geodesic~\cite{pound2008osculating}) as the source to compute the instantaneous self-force
$x^{\mu}(\tau)$ is the particle's location
(in a given system of coordinates $x^{\mu}$)
 and $\tau$ is the particle's proper time.
% on the  original background space-time\MC{Check} and the covariant derivatives are with respect to the original metric as well.
In principle, the $4$-velocity in Eq.\eqref{eq:acc} should correspond to the accelerated orbit  in $g_{\mu\nu}$.
In practise, however,  the $4$-velocity in this accelerated orbit may be replaced with the $4$-velocity of the geodesic (called the  osculating geodesic) in  $g_{\mu\nu}$  which is instantaneously tangential to the accelerated orbit~\cite{pound2008osculating}.
The reason is that the radiation-reaction timescale is much larger than the orbital timescale and the osculating geodesic agrees with the true accelerated orbit to zeroth order for small $h$ (corresponding to small $M_*$ in the case of the tidal force and to small $\mu$ in the case of the self-force).
Therefore, using  one velocity or the other in Eq.\eqref{eq:acc} only changes the force at higher-than-linear order in $h$.
%of the instantaneous geodesic orbit (osculating geodesic) or, equivalently, of the accelerated orbit (it's equivalent because the difference is $O(h)$ and so a higher order contribution)}
When implementing Eq.\eqref{eq:acc} in this paper we shall use this osculating geodesic approximation.
We shall use the symbol $\mathcal{C}$ to denote any  quantity which is conserved along geodesic motion in the  space-time $g_{\mu\nu}$. 
Note that any such quantity  is not necessarily conserved anymore when including the acceleration due to $h_{\mu\nu}$.

%The  background metric on which a small particle is moving is that of a Schwarzschild black hole tidally-perturbed by a remote and slowly-moving (third) body,
%and is  given by the sum of ? Eq.(\ref{eq:eqtide}) together wtih (\ref{eq:electric tensor}).
%With the inclusion of metric perturbations, 
%In this setting, the particle moves on an accelerated trajectory of the original background (Schwarzschild) space-time. Its $4$-acceleration is given by~\cite{Poisson2011}
% \begin{align}
%a^\mu & = -\frac{1}{2} (g^{\mu\nu}+u^\mu u^\nu)(2h_{\nu \lambda;\rho}-h_{\lambda \rho; \nu}) u^\lambda u^\rho,
%\end{align}
%where $u^{\mu}$ is the $4$-velocity of the particle on the  original background space-time\MC{Check} and the covariant derivatives are with respect to the original
%metric as well.
 %, where $g_{\mu\nu}$ is the background metric of the massive black hole
% (which may be, e.g., Schwarzschild or Kerr)
%and $h_{\mu\nu}$ is a gravitational perturbation  (which may be, e.g., a tidal perturbation such as that in Eq.\eqref{eq:eqtide} or the regularized gravitational self-field).

In the second viewpoint, the particle is considered to be following a geodesic of the full  perturbed black hole 
 space-time with  metric  $\Pert{g}_{\mu\nu}\equiv g_{\mu\nu}+h_{\mu\nu}$.
Within the Hamiltonian formalism (see, e.g.,~\cite{vines2015motion,fujita2016hamiltonian} in the context of the conservative self-force), a particle's geodesic motion  in this space-time can be determined by invoking the Hamiltonian equations of motion.
% It is also possible to use the Hamiltonian formalism if $h_{\mu\nu}$ were not static
These equations are
\begin{align}
\frac{d \Pert{q}^\nu}{d\Pert{\tau}} =\frac{\partial H}{\partial \Pert{p}_\nu}\,,\quad \frac{d \Pert{p}_\nu}{d \Pert{\tau}} = - \frac{\partial H}{\partial \Pert{q}^\nu}\,, 
\end{align}
where
% $\Pert{q}^\nu$ are the coordinates of the particle's location, 
 $\Pert{p}_\nu$ is  the  $4$-momentum associated to 
the canonical position $x^{\mu}(\Pert\tau)$ \footnote{In order to not overburden the notation we use the same symbol $x^{\mu}$ to denote the particle's location in the two viewpoints,
although strictly  we  should be differentiating between them since one is a location in $g_{\mu\nu}$ and the other one in $\Pert{g}_{\mu\nu}$ -- it will be obvious from the context which one we mean.}
  and
 $\Pert{\tau}$ is the particles's proper time along the geodesic in $\Pert{g}$. The Hamiltonian is given by 
\begin{align}\label{eq:ham}
H=
\frac{1}{2}\Pert{p}^\mu \Pert{p}^\nu \Pert{g}_{\mu\nu}=
\frac{1}{2}\Pert{p}^\mu \Pert{p}^\nu (g_{\mu\nu}+h_{\mu\nu}).
\end{align}
%As mentioned, we may view the particle as moving on a geodesic of the full, perturbed black hole space-time, $g_{\mu\nu}+h_{\mu\nu}$.
We shall use the symbol $\Pert{\mathcal{C}}$ to denote any  quantity which is conserved along geodesic motion in the perturbed space-time  $\Pert{g}$. 
%Alternatively, we may view the particle to be moving on a space-time with metric $g_{\mu\nu}$ undergoing an acceleration caused by $h_{\mu\nu}$.
%For the sake of simplicity, we shall not try to label the trajectory (position, momentum variables) of particle using different symbols, and we shall state explicitly which picture we adopt for each specific problem. 

We shall essentially adopt the first viewpoint in the calculations from Sec.\ref{sec:conserved quantities} until Sec.\ref{sec:ISCO}, where, for convenience, we shall adopt the second viewpoint.
The rest of this section is organized as follows.
In Sec.\ref{sec:background} we describe geodesic motion on a Schwarzschild black hole background $g_{\mu\nu}$.
In Sec.\ref{sec:tidal field} we give expressions for a tidal perturbation $h_{\mu\nu}$.
Finally, in Sec.\ref{sec:conserved quantities}, we give expressions for rates of change of quantities $\mathcal{C}$ which are conserved along geodesics in $g_{\mu\nu}$.

%---------------------------------------------------------------------------------------------------------------------------------------------------------------------------------------------------------------------

\subsection{Geodesic motion on the black hole space-time} \label{sec:background}

Here we consider geodesic motion on the  black hole background space-time, i.e, with $h_{\mu\nu}=0$ in Eq.\eqref{eq:ham}.

In the case  that the massive black hole is a  Kerr black hole, particles following geodesic motion 
have three conserved quantities: the energy, $E=-u_t$, the component of the angular momentum along the spin ($z$-)axis,  $L_z=u_{\phi}$, and
the  Carter constant $Q$~\cite{carter1968global}. 
Thanks to the three conserved quantities, the radial and angular geodesic motions are separable. 

From now on, however,  we restrict ourselves to the case that the massive black hole is a Schwarzschild black hole.
The Schwarzschild line-element  in Schwarzschild coordinates
$x^{\mu}=\{t,r,\theta,\phi\}$ is
\begin{align}\label{eq:Schw in Schw coords}
ds^2 =
g_{\mu\nu}dx^{\mu}dx^{\nu}=
% -\left ( 1-\frac{2 M}{r}\right) 
-f
 dt^2+
% \frac{dr^2}{1-2M/r}
 \frac{dr^2}{f}
 +r^2 d\Omega^2_2,
\end{align}
where $f\equiv 1-2M/r$ and $d\Omega^2_2=d\theta^2+\sin^2\theta d\phi^2$ is the line-element of the $2$-sphere.
The same metric may be written in Ingoing-Eddington-Finkelstein coordinates as
\begin{align}
ds^2=-
%\left ( 1-\frac{2 M}{r}\right )
f
dv^2+2 dv dr+r^2 d\Omega^2_2\,,
\end{align}
where $v\equiv t+r_*$ and $r_*\equiv r+2M \log \left(1-r/(2M)\right)$ is the tortoise radial coordinate. 

Unlike in Kerr, in the Schwarzschild case the vector angular momentum ${\bf L}$ is conserved, and the particle's motion is planar. 
Because of the additional constraint of planar motion, there are only two effective degrees of freedom left. One of them is radial:
\begin{align}
\left (\frac{d r}{d \lambda} \right )^2 &=
 V_r(r)
\nonumber\\
 &\equiv r^4\left(E^2 -\left ( 1-\frac{2M}{r}\right )\left (1+\frac{L^2_z+Q}{r^2} \right ) \right),
\label{eq:dr/dlambda}
\end{align}
 where we use ``Mino time" $\lambda$, defined via $d\lambda \equiv d\tau/r^2$, to parameterize the trajectory, following the discussion in \cite{drasco2005computing}. 
We note that, in Schwarzschild, the 
Carter constant is given by $Q=u_{\theta}^2+\cot^2\theta L_z^2$ and the
square modulus of the total angular momentum ${\bf L}=(L_x,L_y,L_z)$   is given by
$L^2=||{\bf L}||^2=L^2_z+Q$,
for a given choice of Cartesian coordinates $x$, $y$ and $z$.

 The motion along the $\phi$-direction is given by 
 \begin{align}\label{eq:phimotion}
% \phi =\phi_0+\frac{L_z}{\sin^2\theta} \lambda\,,
 \frac{d\phi}{d\lambda}=\frac{L_z}{\sin^2\theta}.
 \end{align}
%where $\phi_0\equiv \phi (\lambda=0)$.
The motion along the $\theta$-direction (for inclined orbits) can be obtained using a direct mapping from $\phi$ or, alternatively, from
\begin{align}\label{eq:phimotion2}
\left ( \frac{d \theta }{d \lambda}\right )^2 =\frac{L^2_z}{\sin^4\theta}\left ( \frac{d \theta }{d \phi}\right )^2= Q-L^2_z \cot^2\theta\equiv V_\theta(\theta)\,.
\end{align}
The particle moves in the region $\theta\in [\theta_m,\pi-\theta_m]$, with $\theta_m\equiv |\arctan (L_z/\sqrt{Q})|$ being
the angle between ${\bf L}$ and the projection of ${\bf L}$ onto the plane perpendicular to the $z$-axis.

%---------------------------------------------------------------------------------------------------------------------------------------------------------------------------------------------------------------------

\subsection{External tidal field}\label{sec:tidal field}

Poisson and collaborators~\cite{PoissonPRL2005, PhysRevD.81.024029, poisson2015tidal} 
have obtained the metrics of black holes   deformed by  tidal forces which are created by a remote distribution of matter.
They obtain these metrics by solving the perturbative Einstein equation and matching the solution to an external (asymptotic) tidal metric.
In our case, we shall only take into account the leading
%for large $d$. it's leading order because the monopole can be removed by redefining mass and dipole by redifining COM}
 -- quadrupole --  tidal  moment of the field generated by the remote -- third -- body.
This quadrupole moment can be characterized by electric-type tensors $\mathcal{E}_{AB}$, $\mathcal{E}_{A}$ and $\mathcal{E}$, and magnetic-type tensors $\mathcal{B}_{AB}$  and $\mathcal{B}_{A}$, where $A$ and $B$ are indices over the angular degrees of freedom $\theta$ and $\phi$. 
These tensors can be obtained by decomposing the tidal field using tensor, vector and scalar spherical harmonics -- their explicit definitions are given in \cite{PoissonPRL2005,PhysRevD.81.024029,poisson2015tidal}. As the outer object is only  moving slowly, in this paper 
we shall neglect its motion over the orbital timescale of the inner binary. Hence, we shall take the magnetic-type tensor, as well as any derivatives of the electric-type tensor, to be zero in our analysis.
Under these approximations of quadrupole moment and static source, the metric perturbation of a Schwarzschild black hole immersed in an external  tidal field is:

\begin{align}
& h_{vv} = -r^2 f^2 \mathcal{E},\quad  h_{vr} =0, \nonumber \\
%& h_{v A} = -\frac{2}{3} r^3 f (\mathcal{E}^*_A+\mathcal{B}^*_A)\,,\nonumber \\
%& h_{AB} = -\frac{1}{3}r^4\left [ \left ( 1-\frac{2 M^2}{r^2}\right ) \mathcal{E}^*_{AB}+\mathcal{B}^*_{AB}\right ]\,.
& h_{v A} = -\frac{2}{3} r^3 f \mathcal{E}_A\,,\nonumber \\
& h_{AB} = -\frac{1}{3}r^4 \left ( 1-\frac{2 M^2}{r^2}\right ) \mathcal{E}_{AB}\,.
\end{align}
For our purposes, it is more convenient to work with the metric perturbation in Schwarzschild coordinates, which is
\begin{align}\label{eq:eqtide}
& h_{tt} =  -r^2 f^2 \mathcal{E},\quad h_{rr}= -r^2 \mathcal{E},\quad h_{tr}=-r^2 f \mathcal{E}, \nonumber \\
%& h_{tA}= -\frac{2}{3} r^3 f (\mathcal{E}^*_A+\mathcal{B}^*_A)\,, h_{rA}=-\frac{2}{3} r^3  (\mathcal{E}^*_A+\mathcal{B}^*_A)\,,\nonumber \\
%& h_{AB} = -\frac{1}{3}r^4\left [ \left ( 1-\frac{2 M^2}{r^2}\right ) \mathcal{E}^*_{AB}+\mathcal{B}^*_{AB}\right ]\,.
& h_{tA}= -\frac{2}{3} r^3 f \mathcal{E}_A,\quad h_{rA}=-\frac{2}{3} r^3  \mathcal{E}_A\,,\nonumber \\
& h_{AB} = -\frac{1}{3}r^4 \left ( 1-\frac{2 M^2}{r^2}\right ) \mathcal{E}_{AB}\,.
\end{align}

The expressions in~\cite{PoissonPRL2005,PhysRevD.81.024029,poisson2015tidal} for the electric-type (and magnetic-type)
tensors are, in the static limit (within the dynamical timescale of the inner binary), in terms of an external gravitational potential $U_{ext}$, which can be expanded in multipoles. The dipole piece contributes to the
acceleration of the center-of-mass  of the inner binary as studied in~\cite{YunesMiller2011}.
Keeping only the quadrupole order terms, it is, trivially,  
\begin{equation} \label{eq:Uext}
U_{ext}=\frac{M_*(x^2+y^2-2z^2)}{2d^3}=\frac{M_* r^2(1-3\cos^2\theta)}{2d^3}.
\end{equation}
 Here, $z$ is along the direction between the black hole of mass $M$ and the third body  of mass $M_*$, and its origin is at the location of $M$; 
%$x$ and $y$ are the other two  Cartesian coordinates (defined as in flat space-time);
 $\theta$  is the polar angle with respect to the $z$-axis.
 % and the line joining the particle and the mass $M$.
From the expressions in~\cite{PoissonPRL2005,PhysRevD.81.024029,poisson2015tidal} for the electric-type tensors, it then follows that
\begin{align}\label{eq:electric tensor}
&
%\mathcal{E}_{\theta\theta}=\frac{M(3\cos(2\theta)-1)}{2d^3},\quad
%\mathcal{E}_{\phi\phi}=\frac{M\sin^2\theta}{d^3},\nonumber\\ &
\mathcal{E}_{\theta\theta}=-\frac{3M_*\sin^2\theta}{d^3},\quad
\mathcal{E}_{\phi\phi}=\frac{3M_*\sin^4\theta}{d^3},\nonumber\\ &
\mathcal{E}_{\theta}=\frac{3M_*\sin\theta \cos\theta}{d^3},
\quad
\mathcal{E}=\frac{M_*(1-3\cos^2\theta)}{d^3},
\nonumber\\ &
\mathcal{E}_{\theta\phi}=\mathcal{E}_{\phi\theta}=\mathcal{E}_{\phi}=0.
\end{align}

%---------------------------------------------------------------------------------------------------------------------------------------------------------------------------------------------------------------------

\subsection{Changes in ``conserved quantities"}\label{sec:conserved quantities}

Let us now combine a background $g_{\mu\nu}$ and a perturbation $h_{\mu\nu}$ within the first viewpoint  described at the start
of this section.
That is, we consider a particle in accelerated motion due to $h_{\mu\nu}$ on  a background $g_{\mu\nu}$.
%calculate the rate of change of quantities $\mathcal{C}$ 
%Let us now also take into account the perturbation $h_{\mu\nu}$.
%When including acceleration, 
Then,
the  rate of change of a  quantity $\mathcal{C}$, which is conserved along a geodesic in  $g_{\mu\nu}$, may be obtained via
\begin{align}\label{eqchangerate}
\frac{d \mathcal{C}}{d \tau} = &\frac{\partial \mathcal{C}}{\partial p^\nu} \frac{d u^\nu}{d \tau}+\frac{\partial \mathcal{C}}{\partial x^{\nu}} u^{\nu}\, \nonumber \\
= & \frac{\partial \mathcal{C}}{\partial p^\nu} a^\nu\,.
\end{align}
%\MC{here, $a^\nu$ has been defined as $du^{\nu}/d\tau$, but that's different from the true acceleration, which is $a^\mu\equiv D u^{\mu}/d\tau$ as in \eqref{eq:acc}}
%Note that these ``conserved" quantities are defined with respect to  the  and so they may no longer be conserved in the perturbed space-time. 
For example, let us find expressions for the rates of change of the energy, the angular momentum along the $z$-direction and the Carter constant 
in the  Schwarzschild background.
These quantities in the case now of an accelerated orbit are still defined as in the case of a geodesic orbit in Sec.\ref{sec:background}, i.e.,
$E \equiv -p_t$ , $L_z\equiv u_{\phi}$ and
$Q=u_{\theta}^2+\cot^2\theta L_z^2$, respectively.
Here, the $u_{\mu}$ correspond to the accelerated orbit but they are approximated by the values on the osculating geodesic.
Eq.~\eqref{eqchangerate} then yields
\begin{align}\label{eq:dC/dtau}
%p.27WQ
&\frac{d E}{d \tau} =\left(1-\frac{2M}{r}\right)a^t \,,\nonumber \\
&\frac{d  L_z}{d \tau}=r^2 \sin^2\theta a^\phi  \,,\nonumber \\
&\frac{d Q}{d \tau}=2 \cot^2\theta L_z \frac{d  L_z}{d \tau} +2 p_\theta r^2a^\theta\,.
\end{align}
%%As Schwarzschild EMRIs follow planar motion
%For Schwarzschild EMRIs including the gravitational self-force and without an external field, 
%the systems  respect rotational symmetries and so they have no explicit $\phi$-dependence. Since a Schwarzschild geodesic is planar, the
%only degree of freedom that the
% self-force can  depend on is the radial degree of freedom  $r$. Therefore, in order to compute the secular effect
%% (with respect to the orbit of the small body) 
%of the self-force on a small mass moving in Schwarzschild, only averaging over radial motion is needed. On the other hand, 
The presence of a tidal field breaks the spherical symmetry of the background. 
Therefore, the tidal force is generically $\phi$- or $\theta$-dependent as well as $r$-dependent.
As a consequence, its secular effect implies averaging over one of these angular degrees of freedom 
as well as over the radial degree of freedom  $r$.
That is, the Mino-time-averaged Mino-time-derivative of a  quantity $\mathcal{C}$, which is conserved along the osculating geodesic in  $g_{\mu\nu}$, is given by
\begin{align}\label{eq:int1}
\left\langle\frac{d \mathcal{C}}{d \lambda}\right\rangle \equiv \frac{1}{\Lambda_r \Lambda_\theta} \int^{\Lambda_r}_0 d\lambda_r \int^{\Lambda_\theta}_0 d \lambda_\theta \frac{d \mathcal{C}}{d\tau} r^2\,,
\end{align}
or, equivalently, by 
\begin{align}\label{eq:int2}
\left\langle\frac{d \mathcal{C}}{d \lambda} \right\rangle\equiv \frac{L}{2\pi\Lambda_r L_z} \int^{\Lambda_r}_0 d\lambda_r \int^{2\pi}_0 d \phi\, \sin^2\theta\,\frac{d \mathcal{C}}{d\tau} r^2\,.
\end{align}
Here it is understood that in the integrands we write
 $r=r(\lambda_r)$ and, in Eq.\eqref{eq:int1},  $\theta=\theta(\lambda_\theta)$
as functions of Mino time,
as well as $\theta=\theta(\phi)$ in Eq.\eqref{eq:int2}.
The ``Mino time" periods in the $\theta$- and $r$-directions are, respectively,
 $\Lambda_\theta = 2\pi /L$  and 
\begin{align}
\Lambda_r = 2 \int^{\rm r_{max}}_{\rm r_{min}} \frac{dr}{\sqrt{V_r(r)}}\,,
\end{align}
where ${\rm r_{min/max}}$ is the minimum/maximum radius of the orbit.
We note that $\lambda_\theta: 0\to \Lambda_{\theta}$ corresponds to $\phi:0\to 2\pi$.

%---------------------------------------------------------------------------------------------------------------------------------------------------------------------------------------------------------------------
%---------------------------------------------------------------------------------------------------------------------------------------------------------------------------------------------------------------------

\section{Secular effects}\label{sec:secular}

The tidal-induced metric perturbation is stationary in time, which implies conservation of energy (of the orbit on $g_{\mu\nu}$ when including tidal acceleration or, equivalently,
of the geodesic on $\Pert{g}$), i.e.,
the rate of change of the total energy $\Pert{E}\equiv -\Pert{p}_t$ is zero --  instantaneously and so also secularly. The rotational-symmetry along the line connecting the central black hole and the third body also implies conservation of angular momentum along that direction, $z$, i.e., $\Pert{L}_z\equiv \Pert{u}_{\phi}$  is conserved.
% \MC{By switching to the 1st picture, we can also say that the instantaneous rates of change of $E$ and $L_z$ are zero, right? how does that follow from \eqref{eq:dC/dtau}? I don't think the  rates of $E$ and $L_z$ are zero }. 
We note that the relative difference between $\mathcal{C}$ and $\Pert{\mathcal{C}}$ is of order $\mathcal{O}(h)$, 
% Abs. diff.=C-\Pert{C}=O(h^2) => Rel. diff.=(C-\Pert{C})/C=O(h)
which is expected to be small at all times. Therefore, we do not try to highlight their difference when studying secular   evolution of the orbit unless it is necessary.
 
 Now, consider a geodesic orbit in $\Pert{g}_{\mu\nu}$.
 By using the time-reversal symmetry of $\Pert{g}_{\mu\nu}$,
  one can argue that the {\it secular} rate of change of the magnitude of the total angular momentum ${\bf \Pert{L}}$  of this orbit
  % is everything that is said here valid equally  for both  ${\bf L}$ and  ${\bf \Pert{L}}$? which one is the one that precesses? ${\bf \Pert{L}}$  but it only differs from ${\bf L}$ by $O(h)$ }
   must be zero. The argument goes as follows. First, we notice that $\Pert{L}^2$ is a scalar that is invariant under the time-reversal operation. Secondly, because  $g_{\mu\nu}$ and $h_{\mu\nu}$ are independent of time, a time-reversed trajectory still satisfies the correct equation of motion. Based on the above reasoning, if $\Pert{L}^2$ evolves from $\Pert{L}^2_{\rm init}$ to $\Pert{L}^2_{\rm final}$ after some  period of time that is longer  than the orbital  timescale, a time-reversed orbit 
  would evolve $\Pert{L}^2$ from $\Pert{L}^2_{\rm final}$ to $\Pert{L}^2_{\rm init}$. Lastly, it is straightforward to see that 
% a time-reversed orbit can be mapped to its original orbit 
 an orbit is mapped to its  time-reversed orbit
 under 
 the 
 %mirror 
 reflection  operation through a certain symmetry plane.
 The symmetry plane is that formed by the location of $M$, the location of $M_*$ 
 and the point on the orbit 
 %orbit=geod in g+h
   where
% , for example, at places \MC{$\to$ `applied at points on the orbit'?} where 
% irror refl=refl.throught this plane
%palne= this point in eq.21+M+M_*
\begin{align}
\frac{d r}{d \Pert\tau}=0\quad \text{and} \quad \frac{d \theta}{d \Pert\tau}=0\,. 
\end{align}
We argue that an orbit and its reflected one are identical in the sense that the points mapped to each other under reflection should carry the same $\Pert{L}^2$, and consequently, $L^2_{\rm init}$ must be the same 
% if you compare the exact two points then they have the same L 
as $L^2_{\rm final}$.
The joint secular conservation of $\Pert{L}^2=||{\bf \Pert{L}}||^2$ and $\Pert{L}_z$  then means that the opening angle, $\arccos (\Pert{L}_z/\Pert{L})$, between the orbital angular momentum and the symmetry axis of the tidal field must be invariant as well. As a result, the orbital angular momentum ${\bf \Pert{L}}$ can only precess  along the tidal symmetry axis  (this is after orbit-averaging, not instantaneously), with a rate that we compute in Sec.~\ref{sec:Precession}.

Let us now consider another secular effect due to the tidal field. For that purpose, we turn to the viewpoint where the orbit is accelerated in $g_{\mu\nu}$.
We denote by  $\Omega_r\equiv 2\pi/\Lambda_r$, $\Omega_\theta\equiv 2\pi/\Lambda_\theta$ and $\Omega_\phi$
%$\Omega_\phi\equiv 2\pi/\Lambda_\phi$ 
%$\Lambda_\phi$ not defined. check hughes and flanagan paper
the orbital frequencies (with respect to Mino time) associated to, respectively, the $r$-, $\theta$- and $\phi$-motions of a geodesic in $g_{\mu\nu}$.
%Under the osculating orbits approximation \MC{but ~\cite{flanagan2012transient} never talk about osculating orbits ?}, the orbital frequencies of the accelerated orbit can be viewed as evolving along the orbit.
 A ``transient resonance"  is a point on the accelerated orbit such that the radial and angular frequencies
 of the  osculating geodesic at that point
are commensurate with each other: $\Omega_r \colon \Omega_\phi = p \colon q$ (there are only two independent frequencies in Schwarzschild, since the $\theta$- and the $\phi$-dynamics are degenerate and the motion is planar, so we could have equivalently used $\Omega_\theta$ instead of $\Omega_\phi$ in the condition), 
where $p$ and $q$ are prime numbers. In this case, the orbit 
becomes closed and the double integration in Eq.~\eqref{eq:int1} or Eq.~\eqref{eq:int2} reduces to an integral over the closed trajectory. The 
orbital-averaged
rate of change of the magnitude of the angular momentum no longer vanish.
 We evaluate them and discuss their impact on the orbital phase in Sec.~\ref{sec:resonance}.

%---------------------------------------------------------------------------------------------------------------------------------------------------------------------------------------------------------------------

\subsection{Orbital Precession}\label{sec:Precession}

\begin{figure}[tb]
\includegraphics[width=8.4cm]{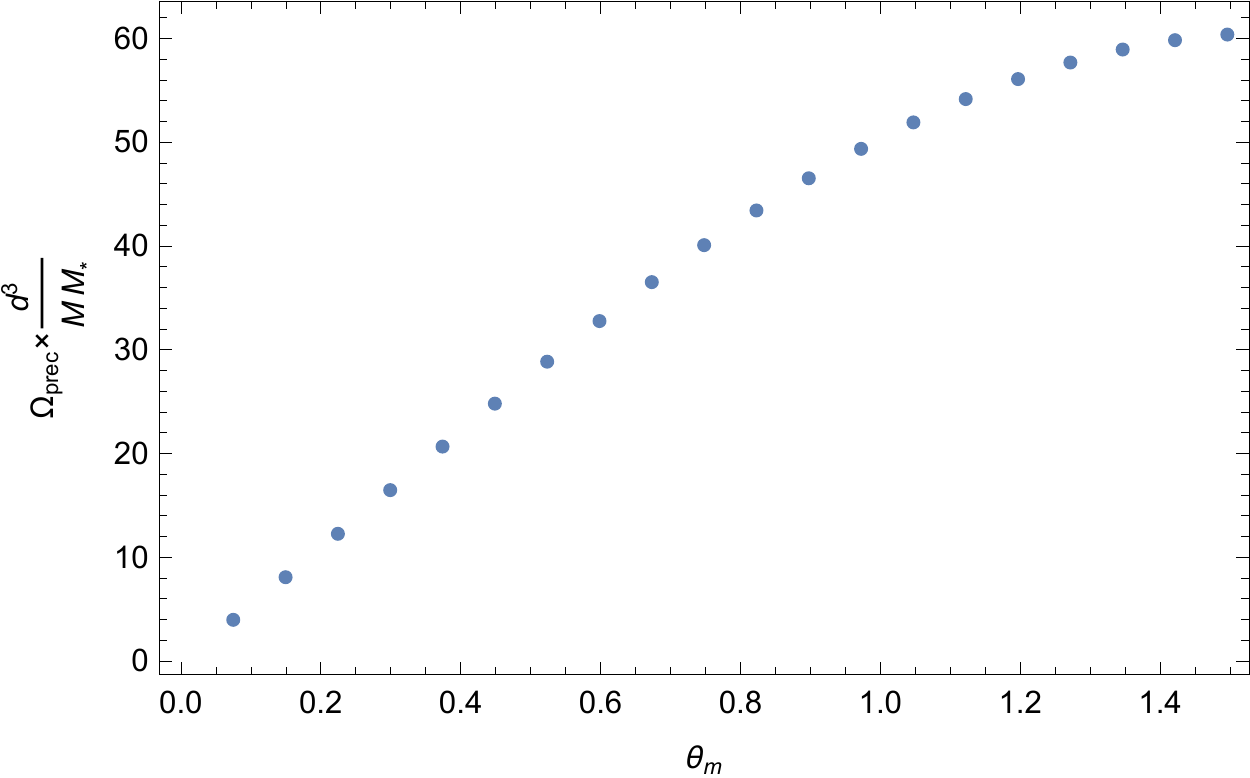}
\includegraphics[width=8.4cm]{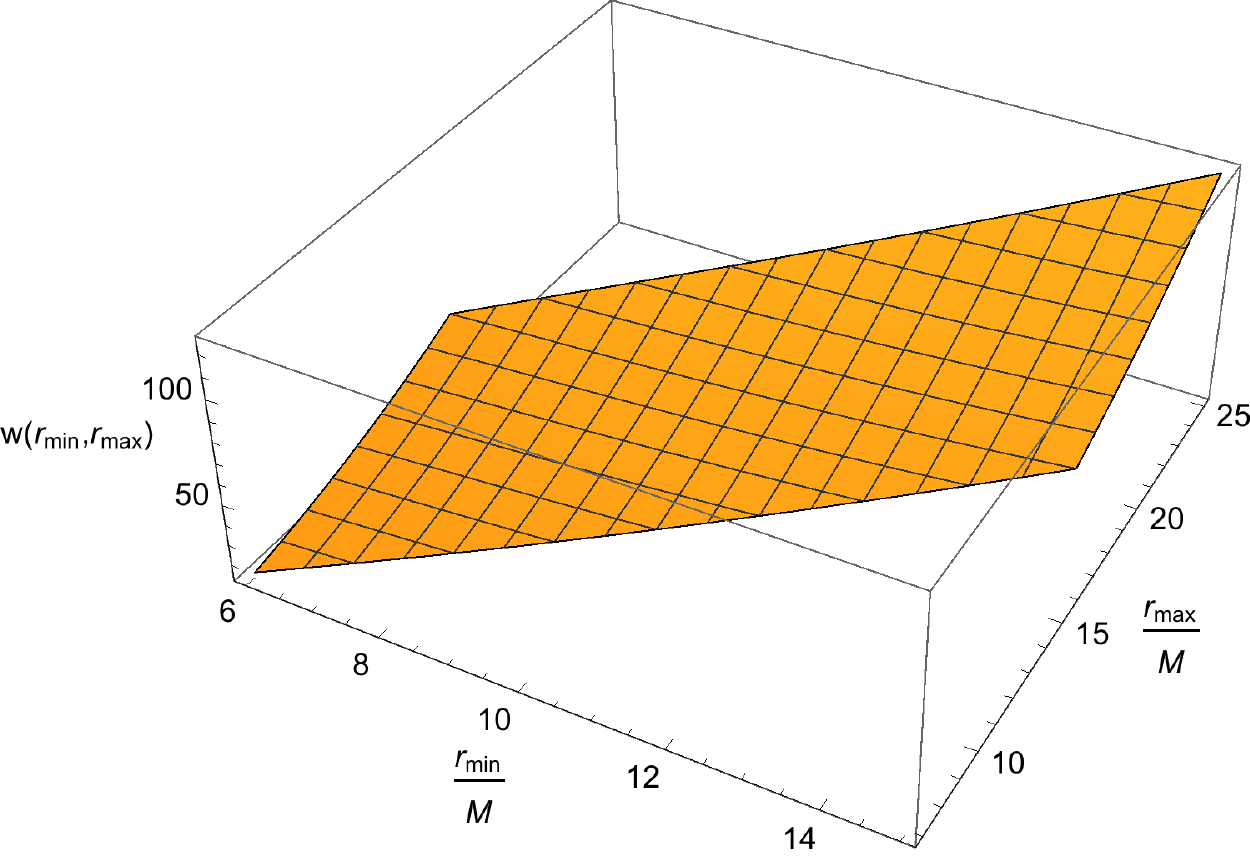}
\caption{Top panel: the precession frequency of the orbital plane as a function of the inclination angle, for the case of $r_{\rm min}=12 M$ and $r_{\rm max}=15 M$. It agrees well with a sine dependence. Bottom panel: $\funcPrec(r_{\rm min},r_{\rm max})$ as defined in Eq.~\eqref{eqpre}.
}
\label{fig:plot1}
\end{figure}

Intuitively speaking, after averaging over the radial and  angular
(either azimuthal or polar) degrees of freedom, the particle trajectory occupies a finite-width ring ($r$ between $r_{\rm min}$ and $r_{\rm max}$)  in the orbital plane of the inner binary.
We note that
when we refer to
 any quantity (such as $U_{ext}$, ${\bf L}$, $\Omega_{\rm prec}$, etc)
 within this subsection, we shall in fact be refering to such orbital-average version of the quantity,
  %that we shall refer to in this subsection will in fact be such orbital-average version of itself, 
  even if we do not say so explicitly.
The mentioned ring has minimum tidal potential energy $U_{ext}$, given in Eq.\eqref{eq:Uext}, if the orbital angular momentum ${\bf L}$ is orthogonal to the tidal symmetry axis, and maximum energy if they are parallel. Therefore, a torque is exerted on the particle orbit, trying to tilt it to the minimum energy state. Such a torque generates precession of the orbital plane, in a similar way to the case of a top precessing under Earth's gravitational field.

In this subsection we adopt the viewpoint of an accelerated orbit in $g_{\mu\nu}$.
In order to evaluate the precession of the orbital plane due to the tidal interaction, we need to compute the secular rate of change of different components of the angular momentum. 
With of the choice of  
the $z$-axis lying along the direction of the central black hole and the third body, $L_z$ must be conserved due to the symmetry argument above. The
%averaged
 precession frequency (with respect to $t$) 
can be computed from the rate of change of $L_x$ and $L_y$ as:
\begin{align}\label{eq:precdef}
%p.22WQ
\Omega_{\rm prec} = \frac{1}{\Gamma_t \Lambda_r \Lambda_\theta} \int^{\Lambda_r}_0 d\lambda_r \int^{\Lambda_\theta}_0 d \lambda_\theta \left(\frac{L_x d L_y}{d\tau} -\frac{L_y d L_x}{d \tau}\right)\frac{r^2}{Q}\,,
\end{align}
%The double integral in \eqref{eq:precdef} is over an osculating geodesic of Schwarzschild, which is not necessarily closed (elliptic)
%The double integral in \eqref{eq:precdef}  should be over  the ccelerated orbit but we can use a geodesic of Schw because the difference is of $O(h)$ and so the difference betweem using one or the other is only at higher order}
where $\Gamma_t$ is the average lapse rate of $t$ with respect to $\lambda$ \cite{drasco2005computing}:
\begin{align}\label{eq:Gamma_t}
%p.28WQ
\Gamma_t = \frac{E}{\Lambda_r}\int^{\Lambda_r}_0  d \lambda \frac{r^2(\lambda)}{1-2M/r(\lambda)}\,.
\end{align} 
We evaluate the quantities in Eqs.\eqref{eq:precdef} and \eqref{eq:Gamma_t} for the osculating geodesic and so, in particular, $E$, $L_z$ and $Q$ are constant.
In order to evaluate the rate of change of $L_x$ and $L_y$, we notice that 
\begin{align}\label{eq:L_x}
%p.24WQ
L_x = -(\sin\phi\, p_\theta+\cot\theta \cos\phi\, L_z)\,.
\end{align}
As a result, its rate of change is related to the acceleration by  (as $L_z$ is conserved)
\begin{align}\label{eq:dL_xdtau}
%p.25WQ
\frac{1}{\mu M}\frac{d L_x}{d \tau} =-a^\theta r^2\sin\phi \,.
\end{align}
The rate of change of $L_y$ can be obtained similarly. 
Separately, one can compute the orbit-averaged  interaction energy (based on the 2-D average of $\tilde{u}_t$) between the ``mass ring" and the tidal field\footnote{Similarly, Hamiltonian 
of the particle on a geodesic of $\Pert g_{\mu\nu}$
 is obtained later in Eq.\eqref{eq:Hamiltonian}.
Such Hamiltonian contains a term that corresponds to the tidal interaction of the ``mass ring".},
which contains a  term 
$E_{\rm int}\propto 1-3(\hat{n} \cdot {\hat{\Pert{\bf L}}})^2 = 1-3\sin^2 \theta_m$ 
 (with a proportionality factor independent of $\theta_m$), where  $\theta_m$
is defined via the last equality as the angle between the tidal symmetry axis $\hat{n}$ (here, parallel to $z$) and the orbital plane. Consequently, the modulus of the torque is 
equal
%p.25WQ
to $d E_{\rm int}/d\theta_m \propto \sin\theta_m \cos\theta_m$. Because the component of the angular momentum orthogonal to $z$ is proportional to $\cos \theta_m$, 
 it must be
 \begin{align}\label{eqpre}
\Omega_{\rm prec} = \funcPrec(r_{\rm min},r_{\rm max}) \frac{M M_*}{d^3}  \sin\theta_m\,,
\end{align}
for a dimensionless function $\funcPrec=\funcPrec(r_{\rm min},r_{\rm max})$;
the proportionality factor $M M_*/d^3$ measures the strength of the tidal-induced acceleration (see Eqs.\eqref{eq:eqtide} and \eqref{eq:acc}).

In Fig.~\ref{fig:plot1}, we present a  calculation of $\Omega_{\rm prec}$, with a normalization constant $d^3/(M M_*)$ to make it dimensionless and to remove the dependence on the strength of the tidal field. 
We have calculated  $\Omega_{\rm prec}$ in the following way. We have used Eq.\eqref{eq:precdef}, with  $L_{x,y}$ and their derivatives calculated via Eqs.\eqref{eq:L_x}
and \eqref{eq:dL_xdtau},
 $a^{\mu}$  via Eq.\eqref{eq:acc}, $h_{\mu\nu}$ from Eq.\eqref{eq:eqtide},
and calculated
$u^{\mu}$   by numerically integrating the geodesic equations  in Schwarzschild.
The (osculating) geodesic in the top panel corresponds to
  $r_{\rm min}=12 M$, $r_{\rm max}=15 M$ and varying values of $Q$ (equivalently, $L_z$ or $\theta_m$).
%However, the $4$-velocity in Eqs.\eqref{eq:L_x}, etc should  be evaluated for the accelerated orbit or, equivalently, for the osculating geodesics. But, presumably, the osculating geodesic changes as the orbit is evolved within one orbital perod, and so the $4$-velocity should correspond to a series of varying osculating geodesic as the the integration over one period is carried out in \eqref{eq:precdef}?
This top panel confirms the dependence on $\sin \theta_m$  given in Eq.\eqref{eqpre}, and the bottom panel gives the numerical value of $\funcPrec(r_{\rm min}, r_{\rm max})$. Apart from  trajectories very close to the MBH, an approximate fit to Fig.~\ref{fig:plot1} is $\Omega_{\rm prec} \sim 1.3 M_* M^{-1/2} r^{3/2}_0/d^3 \sin\theta_m$, with $r_0 =(r_{\rm max}+r_{\rm min})/2$.

The above calculation shows that, in principle, the angular momentum ${\bf \Pert{L}}$ of the inner binary would in principle precess around the direction $\hat{n}$ between the massive black
hole $M$ and the third body $M_*$.
Now, assuming $M \sim M_*$, an order-of-magnitude estimate for the period of the outer binary gives $T_{\rm o} = 2 \pi/\Omega_{\rm o}\sim \left(d^3/M_*\right)^{1/2}$, which is generically much longer than
the precession period: $2\pi/\Omega_{\rm prec} \sim d^3/(M M_*)$. Therefore, we also need to perform an average over the orbit of the third body. This can be done by writing down the equation for the precession of the angular momentum   after averaging over the orbit of the  inner binary, but allowing the direction of the third body ($\hat n$) to be time-dependent.
From Eq.\eqref{eqpre} (for simplicity, here we do not distinguish between ${\bf L}$ and ${\bf \Pert{L}}$),
\begin{align}\label{eq:dL/dt}
\frac{d {\bf \Pert{L}}}{d t} = \funcPrec(r_{\rm min},r_{\rm max})  \frac{M M_*}{d^3} (\hat{n} \cdot{\bf  \hat{\Pert{L}} }) \hat{n} \times {\bf \Pert{L}}\,.
\end{align}

Let us assume that the motion of the third body is on some arbitrary $x'$--$y'$ plane, so that we can write $\hat{n}=\sin (\Omega_{\rm o} t)\, \hat{x}'+\cos (\Omega_{\rm o} t)\, \hat{y}' $. 
The angular momentum of the outer binary is therefore perpendicular to the $x'$--$y'$ plane and so parallel to the $z'$ axis.
By plugging this expression for $\hat{n}$ into the above equation and averaging  over an orbital period of the outer binary, $2 \pi/\Omega_{\rm o}$, we obtain
\begin{align}\label{eq:vg dLdt}
\left \langle \frac{d {\bf \Pert{L}}}{d t} \right \rangle_{\rm o} = -\funcPrec(r_{\rm min},r_{\rm max})  \frac{M M_*}{2 d^3} (\hat{z}' \cdot {\bf  \hat{\Pert{L}} }) \hat{z}' \times {\bf \Pert{L}}\,.
\end{align}
Thus, now ${\bf \Pert{L}}$  precesses around $z'$: Fig.\ref{fig:system}.
Physically Eq.~\eqref{eq:dL/dt} and \eqref{eq:vg dLdt} describes the precession generated by the quadrupole moment-curvature coupling of the inner binary.
For the MBH (outer) binary scenario considered here, the precession period is generically longer than the LISA observation timescale.  
However, we note that the precession effect also extends to the Newtonian regime as well as to comparable-mass binaries (instead of EMRIs). 
Thus, let us consider here --and only here-- the case of stellar-mass BH binaries  close to    a MBH of mass $M_*$, which could be  relevant
sources for both LISA and LIGO detections \cite{antonini2012secular,thompson2011accelerating,antonini2014black,silsbee2016lidov,vanlandingham2016role}.
In this case, the precession period can be estimated as
\begin{align}
\frac{2\pi}{\Omega_{\rm prec}}& \sim \frac{2 \pi d^3 M^{1/2}}{ 1.3 \sin\theta_m r^{3/2}_0 M_*} \nonumber \\
&\sim 2.6 \, {\rm day} \left ( \frac{d}{30 M_*} \right )^3 \left ( \frac{M_*}{M_{\rm SgA^*}}\right )^{2} \left (\frac{f_{\rm GW}}{1 {\rm mHz}} \right )^{2/3},
\end{align}
where $\theta_m$ is taken to be $\pi/4$ for illustration purposes, $f_{\rm GW}$ is the GW frequency (twice the orbital frequency) of the stellar-mass binary and the component masses are assumed to be $10 M_\odot-10 M_\odot$. Notice that such binaries (as well as EMRIs ) are likely to be eccentric due to the KL mechanism. Therefore, the waveform also contains a frequency component $\sim f_{\rm GW} (1-e)^{-3/2}$ (where $e$ is the eccentricity) corresponding to the pericenter passage.

%---------------------------------------------------------------------------------------------------------------------------------------------------------------------------------------------------------------------

\subsection{Resonance}\label{sec:resonance}

In this subsection we consider a point in an accelerated orbit of the particle where the osculating geodesic (in $g_{\mu\nu}$) is a resonant point.
At a resonance, the osculating orbit is closed and we no longer consider ``phase-space-averaged"-orbits which span a two-dimensional ring on a plane. In this sense, this situation is more similar to the Newtonian limit, which may be viewed as a $\Omega_r \colon \Omega_\phi=1 \colon 1$ resonance. In the Newtonian limit, the orbital eccentricity can be boosted to very high values via the Kozai-Lidov mechanism \cite{kozai1962secular, lidov1962evolution}. Following the above analogy, we also expect a non-trivial change of eccentricity and angular momentum of the relativistic orbit during a resonance phase. In particular, the total angular momentum might be boosted, as compared to the monotonic  reduction generated by the dissipative self-force \footnote{Note that the dissipative self-force may increase the eccentricity below a certain critical radius \cite{PhysRevD.47.5376}.}.
%\MC{That's not quite true, as the dissipative SF may increase the eccentricity below a certain critical radius --  see, eg, PRD 47, 5376 (1993)}.

In the calculation of the precession, an order-of-magnitude analysis showed that we could not neglect the orbit of the outer binary when $M\sim M_*$.
Let us carry out a similar order-of-magnitude analysis here.
% For inner binary orbit, $T_{\em o} \sim M (r_0/M)^{3/2}$ is of the same order as $M$.
Also, the timescale of a transient resonance driven by the dissipative self-force generally scales as $T_{\rm res} \sim \mu^{-1/2}M$. By comparing it to the orbital timescale of the third body, we have $T_{\rm res}/T_{\rm o} \sim \left(a_{\rm tide}/a_{\rm s}\right)^{1/2} (r_0/M)^{-11/4}  \ll 1$. Therefore, for the case we study here, the static approximation for the tidal field applies. 

In this subsection, we continue to choose the $z$-axis to be parallel to the symmetry axis of the tidal field. Such setup is slightly different from the celestial coordinate setting in previous studies of hierarchical triple system in Newtonian and post-Newtonian regimes \cite{lithwick2011eccentric,li2015implications,naoz2013resonant}, as we do not perform the average over the third body's orbit. On the other hand, our coordinate choice ensures rotational symmetry of the space-time around the $z$-axis, so that $L_z$
%We could equally write $\Pert{L}_z$ instead: $L_z$ differs from $\Pert{L}_z$ by at most $O(h)$, and so it differs from its initial value by $O(h)$ and that doesn't affect the resonance effect to the leading order we're using 
 must be conserved. 

Suppose that $\Omega_r \colon \Omega_\phi = p \colon q$, where $p$ and $q$ are coprime numbers. Then the integration 
of the rate of change of a quantity $\mathcal{C}$ over a resonant closed orbit  is

\begin{align}\label{eq:avg C resonance}
\left\langle \frac{d \mathcal{C}}{d \lambda} \right\rangle_r\equiv \frac{1}{\Lambda } \int^{\Lambda}_0 d\lambda \frac{d \mathcal{C}}{d\tau} r^2\,,
\end{align}
where $\Lambda\equiv p \Lambda_r =q \Lambda_\theta$. Such an integration is independent of  the longitude of the ascending node for generic inclined orbits \cite{naoz2013secular}, but it does depend on the integration constant $\lambda_{\theta 0}$  in the $\theta$-motion, 
\begin{align}
\lambda_\theta=\lambda_{\theta 0}+\int^\theta_{\theta_m} \frac{d\theta}{\sqrt{V_{\theta}(\theta)}}\,,
\end{align}
which is related to the argument of the periastron \cite{naoz2013secular}
% The  argument of the periastron might not be exactly equal to $\lambda_{\theta 0}/\Lambda_\theta$, there might be a proportionality factor or something, not sure

In order to illustrate this point, we pick a resonance point with $\Omega_r/\Omega_\phi = 1/2$. This can be achieved with a one-parameter family of radius (e.g., either $r_{min}$ or $r_{max}$). For convenience, we choose  $r_{min}=7M$ and $r_{max}=9.39117M$, although this choice is not unique. 
Also, we choose the inclination angle to be $\theta_m=\pi/4$ ($\sin \theta_m =\hat{z}\cdot \hat{L}$).
%, which could be chosen to be other values as well. 
From Eqs.\eqref{eq:avg C resonance} and \eqref{eq:dC/dtau}, with the tidal acceleration from Eqs.\eqref{eq:acc} and \eqref{eq:eqtide}, we calculated
the orbital-averaged rate of change of the Carter constant $Q=L^2-L^2_z$ as a function of
%the argument of the periastron, 
$\lambda_{\theta 0}/\Lambda_\theta$, for this chosen resonant orbit.
We present this rate of change (which is trivially related to the corresponding rate of change of the total angular momentum $L$) in Fig.\ref{fig:plot3}.
This plot clearly shows that the averaged rate of change of the Carter constant or, equivalently, of the total angular momentum, is nonzero during a resonance.
 In addition, the dependence on  
$\lambda_{\theta_0}$ is well described by a sinusoidal function $\sin(4 \pi  \lambda_{\theta_0}/ \Lambda_\theta)$, as the pattern of the closed orbits repeats itself  every
% Approximately? probably related to symmetry of tidal field
$180$-degree rotation in the argument of the periastron. 

\begin{figure}[tb]
\includegraphics[width=8.4cm]{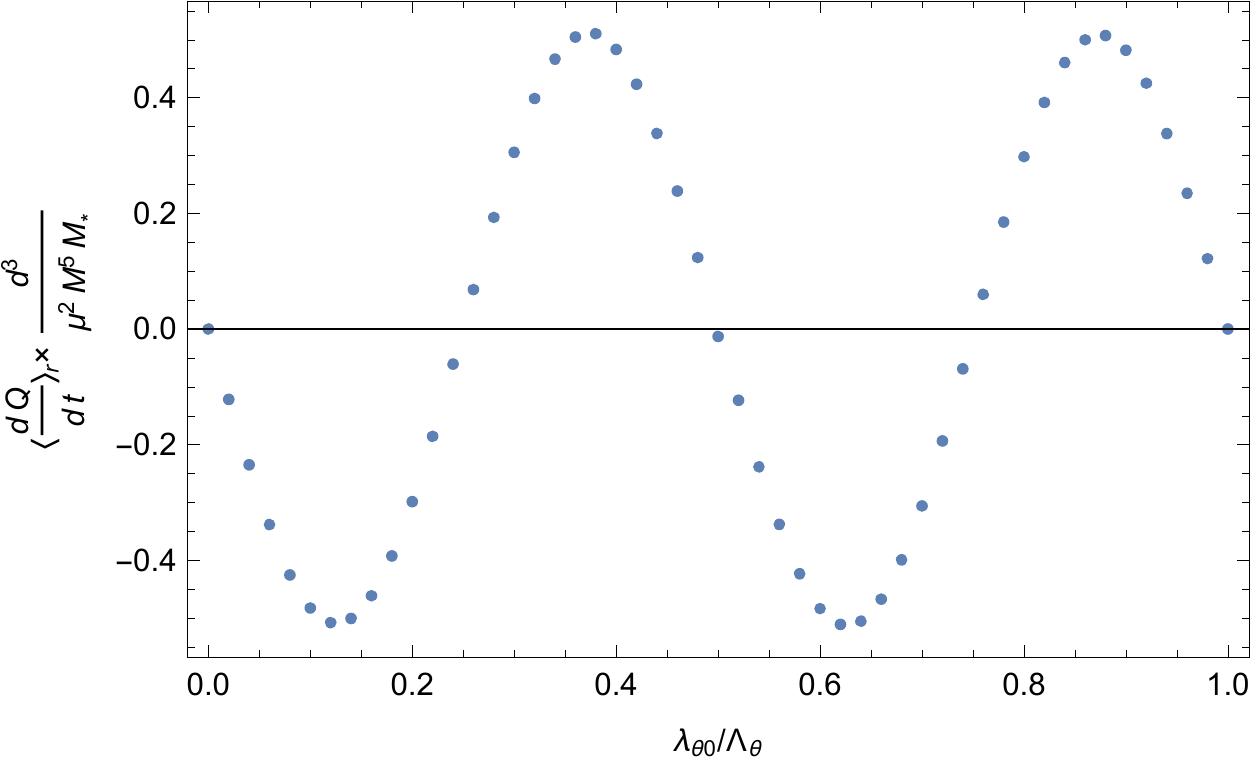}
\caption{The resonant-orbit-averaged
rate of change of the Carter constant $Q=L^2-L^2_z$ as a function of the argument of the periastron, $\lambda_{\theta 0}/\Lambda_\theta$. The equivalent rates of change of $E$
 and $L_z$ are both zero.}
\label{fig:plot3}
\end{figure}

We performed a similar calculation of the averaged rates of change $E$ and $L_z$ and found that, as opposed to the calculation for $Q$,
they are both zero (within the prescribed numerical accuracy of our calculation).
%After applying the average over  a resonant orbit, We find (numerically using Eq.\eqref{eq:dC/dtau}) that the average change rate of $E$ and $L_z$ are both zero, but the rate of change of Carter constant or total angular momentum is nonzero during resonance, as shown by Fig.~\ref{fig:plot3}.
The fact that  the energy is conserved during a resonance phase due to the conservation of $\tilde{E}$ in the perturbed spacetime, and is consistent with previous studies in the Newtonian regime \cite{naoz2013secular}, although in that case  averaging over the orbital phase has been applied to the third body, which is likely to have a longer period than the resonance crossing time $\sim \mu^{-1/2}M$ in the systems that we are considering.

 In general, $L$ might be boosted at a resonance, as opposed to the monotonic  reduction generated by the dissipative self-force \footnote{Note that the dissipative self-force may increase the eccentricity below a certain critical radius \cite{PhysRevD.47.5376}.}.
This resonance effect  due to the tidal force is in stark contrast with the case of the conservative self-force, which cannot drive resonances
since it does not have an explicit $\phi$-dependence~\cite{flanagan2012transient}. Its existence also implies that the Newtonian KL effect and this relativistic KL effect might be formulated in a more general resonance kinetic theory framework. 
As Schwarzschild EMRIs are driven only by the dissipative  self-force during a resonance, the shift of conserved quantities is $\sim\mu^{1/2}$ and the resulting phase modification during the radiation-reaction timescale ($\sim M/\mu$) is $\sim\mu^{-1/2}$, which is much larger than unity. We have shown that tidal interaction also drives the evolution of conserved quantities during resonance. Its contribution to  the  phase error in the radiation-reaction timescale is $\sim \mu^{-1/2}\, a_{\rm tide}/a_{\rm s}$. We emphasize that the secular amplification of order $\mu^{-1/2}$ accumulates over the transient resonance scale. Therefore, this is an effect which cannot be captured by directly evaluating $a_{\rm tide}/a_{\rm s}$ in the dynamical regime as in~\cite{YunesMiller2011}.

%{\bf Here Marc please add the plots for waveform and corresponding phase variation due to tidal interaction.}

%---------------------------------------------------------------------------------------------------------------------------------------------------------------------------------------------------------------------
%---------------------------------------------------------------------------------------------------------------------------------------------------------------------------------------------------------------------

\section{ISCO shift}\label{sec:ISCO}

As is well-known, the conservative piece of the gravitational self-force on the smaller mass in an EMRI system causes a shift in the ISCO frequency and radius,
with respect to the test particle case~\cite{BarackSago09,isoyama2014gravitational}.
Similarly, in a  hierarchical three-body system, the tidal field by the third body also modifies the ISCO frequency and radius of a Schwarzschild EMRI. 
In this section, we directly derive the shift in the ISCO frequency,  radius, energy and angular momentum due to the tidal field,  to leading order in $\epsilon \equiv M^2 M_*/d^3$. 
Therefore, while so far we have been considering general inspiral orbits, we now consider orbits which would be circular in the absence of the tidal field
(similarly to the ISCO shift in the gravitational self-force case, where the  orbits considered are those which would be circular in the absence of the dissipative self-force). 
%We note, however, that the tidal field breaks the (helical) symmetry and so there exist no circular orbits in its presence (as opposed to the conservative self-force case, where radial orbits do exist).
%Therefore, we shall consider an orbital-average version of a circular orbit, as described later.

In this section, for the convenience of analysis and to allow easier implementation of previous results in the gravitational self-force problem, we choose a new coordinate system
%\MC{Does this simplify any eq. in the paper? if not, we could just say that the calculation is done more easily in this way but not change the meaning of $z$ within the paper? yes it simplifies eg \eqref{eq:avg Ham}}
 such that the $z$-axis is orthogonal to the orbital plane of the inner binary (while it still goes through $M$). We shall also adopt the viewpoint that the particle is moving in a geodesic of the perturbed Schwarzschild space-time, similar to the treatment in \cite{detweiler2008consequence,isoyama2014gravitational}. Correspondingly, the angles $\theta$ and $\phi$ are now the polar
 and azimuthal angle, respectively, with respect to this new $z$-axis.
Thus, in particular, 
the instantaneous angular velocity of the particle
% the orbital frequency
 is given by $\Omega\equiv d \phi/dt$.

Notice that the quantities for which we compute the ISCO shift are all gauge-dependent quantities (e.g., $\Pert{E}$ is conserved but  gauge-dependent). Therefore, any result obtained here has to be associated with the gauge that we have chosen. This observation and the associated ambiguity has been emphasized in \cite{barack2001gravitational} in the context of the gravitational self-force problem. In that context, this issue is partly resolved by Detweiler in~\cite{detweiler2008consequence} by considering 
quantities  (such as $\Omega$) which, on ``quasi-circular" orbits
% should we define them?
are
pseudo-invariants with respect to helical-symmetric gauge transformations:
\begin{align}\label{eq:helical symm}
\mathcal{L}_k g_{\alpha \beta}=0,\quad \mathcal{L}_k h_{\alpha \beta} =\mathcal{O}(\mu^2)\, ,
\end{align}
where $\mathcal{L}_k$ is the Lie-derivative with respect to the helical symmetry vector $k=k^a\partial_{x^a} =\partial_t+\Omega\partial_\phi$. In our case, however, the external field itself breaks such symmetry, and it is not clear whether there is a similar construction of pseudo-invariant quantities.
%Notice that in the intermediate steps we shall compute shift of quantities such as ISCO radius and etc., which are gauge dependent. However, the final value for frequency modification should be a gauge-independent quantity by definition, although this is not explicit from the derivation. 
On the other hand, it is possible to assert an angular-averaged version of helical symmetry:
\begin{align}\label{eq:angavg}
\int^{2\pi}_0 d\phi\, \mathcal{L}_k h_{\alpha \beta} =\mathcal{O}(\mu^2)\,.
\end{align}
If the above requirement is satisfied, it is straight-forward to modify Detweiler's derivation of gauge invariance of $\Omega$ 
%(represented by $\Omega$  in \cite{detweiler2008consequence}) 
to prove the invariance of $\int^{2\pi}_0 d\phi \,\Omega$ on ``quasi-circular" orbits with respect to any gauge choices satisfying Eq.~\eqref{eq:angavg}. Note that, based on Eqs.~\eqref{eq:phimotion} and \eqref{eq:phimotion2}, the average over $\phi$ can be replaced by an average over a period of $\lambda_\theta$. 

There is one more restriction on the gauge choice, though a rather natural one.
Detweiler requires the gravitational perturbation not only to respect the helical symmetry Eq.\eqref{eq:helical symm} but also the reflection symmetry through the equatorial plane.
The tidal field in our system, however, leads to the violation of this symmetry. If one does not require reflection symmetry, the changes in the metric perturbation under a gauge
transformation are then given by those in Eqs.B2-B7~\cite{detweiler2008consequence} with the only following modification:
\begin{equation}\label{eq:gauge}
\Delta h_{\phi \phi} \to \Delta h_{\phi \phi}-\xi^{\theta} 2 r^2 \sin \theta \cos \theta,
\end{equation} 
where $\xi^{\mu}$ is the gauge vector.
Because here we are considering orbits on the equatorial plane (where  $\cos \theta=0$), the 
modification term in Eq.\eqref{eq:gauge} vanishes as long as $\xi^{\theta}$ is finite. As a result, the gauge-invariance of  $\int^{2\pi}_0 d\phi \,\Omega$  is mantained
 under gauge transformations that preserve Eq.\eqref{eq:angavg} and with $\xi^{\theta}$ finite.

Because the tidal field breaks the axi-symmetry  when the orbital angular momentum is not aligned with the symmetry axis of the tidal field, there is no innermost orbit with strictly circular motion (in fact, there are no strictly-circular orbits at all). Instead, the true trajectory $\gamma'$ (on the full metric $\Pert{g}_{\mu\nu}$) has a slight  oscillation in the radial coordinate of magnitude $\delta r \sim \epsilon M$. In fact, if for a moment we take the point of view that the particle is moving in an accelerated orbit in Schwarzschild space-time, the tidal forces (proportional to $\epsilon$) in both radial and azimuthal angle directions contain pieces that are periodic in $\phi$ and pieces that are independent of $\phi$. 
By solving the equations of motion including the tidal acceleration, it is easy to see that
the radial motion of the ISCO orbit (note that the orbit is not actually circular, but it is an innermost stable, circular {\it mean} orbit) can be written as 
%$r=r_{\rm mean} +\mathcal{\epsilon} \{ \sin\phi, \cos \phi, \sin 2\phi, \cos 2 \phi\}$ ( $\epsilon$ times a linear combination of the terms in the brackets with coefficients that are independent of $\phi$),
$r=r_{\rm mean} +\mathcal{\epsilon} \left(c_1 \sin\phi+c_2 \cos \phi+c_3 \sin 2\phi+c_4 \cos 2 \phi\right)$  for some $r_{\rm mean}$, 
where $c_i$ ($i={1,2,3,4}$) are coefficients independent of $\phi$.
Such description should also be valid in the perturbed space-time  picture to which we now return, i.e., the orbit is closed to leading order  in $\epsilon$.
%As a result, the Given a Hamiltonian
%$\Hdim$ of the EMRI+tidal interaction system, the requirements for the ISCO can be translated into requirements over the phase-space averaged motion:
%\begin{align}
%& \left .\int^{\Lambda_\theta}_0 d\lambda_\theta  \int^{\Lambda_r}_0 d \lambda_r \,\Hadim \right |_{\gamma'}=\frac{1}{2}\,,\label{eq:EMRI cond1}\\
%& \left . \int^{\Lambda_\theta}_0 d\lambda_\theta \int^{\Lambda_r}_0  d \lambda_r \, \frac{d p_r}{d\lambda} \right |_{\gamma'} =0\,,\\
%&  \left . \frac{\partial}{\partial r_{\rm mean}}\int^{\Lambda_\theta}_0 d\lambda_\theta  \int^{\Lambda_r}_0   d \lambda_r \,\frac{d p_r}{d\lambda}  \right |_{\gamma'}  =0\,,
%\end{align} 
%\MC{Where do you get \eqref{eq:EMRI cond1} from?}
%where $d p_r/d\lambda$ is to be replaced by `$-r^2\partial \Hdim/\partial r$' according to the Hamiltonian equations of motion, and 
Given a Hamiltonian
$\Hdim$ of the ``EMRI+tidal interaction system", we define $\Hadim\equiv \Hdim /(\mu^2M^2)$
 as a dimensionless quantity.
 Let us denote by  $\gamma$ the ``mean" circular orbit (on $\Pert{g}$) with radius $r_{\rm mean}$.
 We notice that $\gamma$  is not strictly geodesic and that it has a Hamiltonian order $\mathcal{O}(\epsilon^2)$ 
% see above Eq.2~\cite{isoyama2014gravitational}
 away from the true trajectory \footnote{A similar treatment was employed in~\cite{isoyama2014gravitational} to compute the ISCO shift on the equator of Kerr due to the gravitational self-force.}. In other words, the effect of radial motion only contributes with $\mathcal{O}(\epsilon^2)$ terms  to the Hamiltonian. 
 As a result, we can replace $\gamma'$ with the mean circular trajectory $\gamma$, which is  convenient for practical calculations.
Accordingly, we calculate $\HadimAvg$, where we define the ISCO-orbital-average of a quantity $\mathcal{A}$ as
  \begin{equation}  \label{eq:avg Ham}
%  \HadimAvg
  \AvgI{\mathcal{A}}
   \equiv \frac{1}{2 \pi}\int^{2\pi}_0d\phi %\left.\Hadim\right|_{\gamma},
    \left.\mathcal{A}\right|_{\gamma},
        \end{equation}
where 
%$\left.\Hadim\right|_{\gamma}$ is $\Hadim$ evaluated along $\gamma$.
$\left.\mathcal{A}\right|_{\gamma}$ is $\mathcal{A}$ evaluated along $\gamma$.
In the case of $\left.\Hadim\right|_{\gamma}$, we obtain it
 from Eq.\eqref{eq:ham}, using Eqs.\eqref{eq:eqtide} and \eqref{eq:electric tensor} and setting $\Pert p^r=0$, $\Pert p^\theta=0$ and $\theta=\pi/2$ (where $\theta$ is with
respect to the new $z$-axis).
% and after averaging over $\lambda_\theta$.
The  resulting, dimensionless and averaged, Hamiltonian is
\begin{align}
&
\HadimAvg = -\frac{\Pert E^2}{2 (1-2M/r)}+\frac{\Pert L^2}{2r^2}-
\nonumber \\ &
\frac{M_* \left(1-3(\hat{n}\cdot \hat{\Pert{\bf L}})^2\right)}{4 d^3}
%\nonumber \\ &
%\times 
\left(\Pert E^2 r^2+\left(1-\frac{2M^2}{r^2}\right)\Pert L^2\right)\,.
\label{eq:Hamiltonian}
\end{align}
% inner prod. $\hat{z}\cdot \hat{L}$ is indep. of coord. syst.

The ISCO condition   now reduces to (without distinguishing $r$ from $r_{\rm mean}$ here and by adopting the argument in \cite{isoyama2014gravitational}):
\begin{align}\label{eqsimreq}
& \HadimAvg =-\frac{1}{2}\,,\nonumber \\
&\frac{\partial \HadimAvg}{\partial r}=0\,,\nonumber\\
& \frac{\partial^2 \HadimAvg}{\partial r^2} =0\,.
\end{align} 
Let us define $\eta \equiv \epsilon \left(1-3(\hat{n} \cdot \hat{\Pert{\bf L}})^2\right)/4$. Then, 
when including the tidal field, the
energy, angular momentum, radius and orbital frequency at the mean orbit of ISCO  are, given by, respectively,
%-induced ISCO shift \MC{but it's not a shift that is being expanded in \eqref{eq:ISCO shift}? right, so rephrase} of radius and conserved quantities can be expanded in $\eta$ as:

\begin{align}
&
% \frac{E}{\mu M}
\Pert{E}
=\Pert E_0+\eta\ \Pert E_1+\mathcal{O}(\eta^2)\,,\nonumber \\
& \frac{\Pert L}{M}=\Pert L_0+ \eta\ \Pert L_1+\mathcal{O}(\eta^2)\,, \nonumber \\
& \frac{r}{M} = r_0+\eta\ r_1+\mathcal{O}(\eta^2)\,,\nonumber \\
& M\AvgI{\Omega}  = \Omega_0 +\eta\ \Omega_1+\mathcal{O}(\eta^2)\, ,
\label{eq:ISCO shift}
\end{align}
where $r_0=6$, $\Pert{E}_0 =\sqrt{8}/3$, $\Pert{L}_0=2\sqrt{3}$ and $\Omega_{0}=1/(6\sqrt{6})$ are  the values for a test particle on the ISCO in Schwarzschild 
and  $r_1$, $E_1$, $L_1$ and $\Omega_{1}$ are defined with respect to their expansion order in $\eta$. By plugging Eq.\eqref{eq:ISCO shift} into Eq.~\eqref{eqsimreq} we  obtain the
following shifts:
\begin{align}
%r_1=-3072 ,\quad E_1=-\frac{152}{3},\quad L_1=174\sqrt{6} \,.
r_1=3072 ,\quad E_1=-\frac{152\sqrt{2}}{3},\quad L_1=-348\sqrt{3} \,.
\end{align}

The ISCO frequency in the perturbed space-time is given by ~\cite{detweiler2008consequence} and we apply it on the mean circular orbit of ISCO (i.e., $\gamma$):
\begin{align}\label{eq:Omega ISCO}
\AvgI{\Omega}  =&\AvgI{ \frac{d \phi}{dt}} =\AvgI{ \frac{u^{\phi}}{u^t}} \nonumber \\
=& \frac{M}{r^3}-\frac{r-3M}{2 r^2} \tilde{u}^{\mu} \tilde{u}^{\nu}\AvgI{ \partial_r h_{\mu\nu}}\,.
\end{align}
%where any quantity within angle brackets is defined  as in Eq.\eqref{eq:avg Ham} but with the quantity in question in the place of $\Hadim$.

In order to evaluate this expression, we need the following orbital-averages,
% $ \AvgI{h_{tt}}$, $\AvgI{ h_{t\phi} }$ and $\AvgI{ h_{\phi\phi}}$. These values
which
 are readily obtained:
\begin{align}\label{eq:avg pert}
& \AvgI{h_{tt}} = \frac{M_* r^2 \left(1-3(\hat{n} \cdot \hat{\Pert{\bf L}})^2\right)}{2 d^3}\left (1- \frac{2M}{r}\right )^2, \nonumber \\
& \AvgI{h_{t \phi}} =\frac{M_* r \left(1-3(\hat{n} \cdot \hat{\Pert{\bf L}})^2\right)}{2 d^3}\left (1- \frac{2M}{r}\right ), \nonumber \\
& \AvgI{h_{\phi\phi}} = r^2 \AvgI{{h}_{tt}}\,.
\end{align}

From Eqs.\eqref{eq:Omega ISCO}, \eqref{eq:ISCO shift} and \eqref{eq:avg pert}, we finally obtain
\begin{align}
 \Omega_1 = 
 %\frac{ L_1}{r_0^2}-2\frac{L_0  r_1}{r^3_0}=\eta \frac{599}{3\sqrt{6}} \,.
-\frac{277 }{54}
\end{align}
for the shift in the ISCO frequency.
%{\bf Here are the $h_{tt}, h_{\phi\phi}$ on the equatorial plane, after averaging over $\phi$ from $0 \to 2\pi$}
%\begin{align}
%& \bar{h}_{tt} = \frac{M_* r^2 [1-3(\hat{n} \cdot \hat{L})^2]}{2 d^3}\left (1- \frac{2M}{r}\right )^2 \nonumber \\
%& \bar{h}_{\phi\phi} = r^2 \bar{h}_{tt}\,.
%\end{align}
Therefore, the tidal field could give rise to either a positive or a negative shift in the ISCO frequency, depending on the sign of $\eta$. Inclined orbits with $\hat{n} \cdot {\hat{\Pert{\bf L}}}=1/\sqrt{3}$ have no tidal-induced shift in the ISCO frequency.

%---------------------------------------------------------------------------------------------------------------------------------------------------------------------------------------------------------------------
%---------------------------------------------------------------------------------------------------------------------------------------------------------------------------------------------------------------------

\section{Discussion and Conclusion} \label{sec:conclusions}

We have performed an analysis of the general-relativistic dynamics of a Schwarzschild EMRI (without gravitational self-force) residing in an external quadrupole tidal field. As discussed earlier, the detection rates of such systems are still subject to uncertainties in EMRI merger rate as well as the merger history of MBHs before GW 
radiation takes over. This also means that a possible detection of such event would also shed light on the myth of the MBH merger mechanism. It would also provide a unique opportunity to test a perturbed Schwarzschild/Kerr metric predicted by General Relativity, as it has distinctive dynamic and waveform features compared to isolated EMRI systems.

We have discussed three interesting relativistic effect due to the tidal interactions.
First, in the non-resonant phase of the EMRI orbit, the main secular effect of this tidal interaction is the precession of the  orbital plane around the orbital angular momentum of the outer binary. This precession may contribute at order $\mathcal{O}(2\pi)$ to the phase of the waveform during the precession timescale, given by  $\sim M/\Omega_{prec}\sim M/\epsilon$}. However, such precession timescale for EMRIs might be longer than observation timescale of LISA, whereas a similar mechanism applied to stellar mass binary systems near a SMBH gives $\mathcal{O}(\rm days)$ precession timescale in the LISA band.
Second, during the resonant phase, 
%on the other hand, 
the magnitude of  the angular momentum may increase or decrease, in stark comparison with the monotonic suppression driven by the dissipative part of the self-force.
The  fractional change in the magnitude of the angular momentum driven by the tidal-field during a resonance is, when including the dissipative self-force as well as the tidal force,
 proportional to $\mu^{1/2}a_{\rm tide}/a_s$, and the resulting orbital phase modification is $\sim \mu^{-1/2}a_{\rm tide}/a_s $. This value could be greater than phase resolution of LISA depending on the strength of the tidal field and the parameters of the inner binary. 
Finally, in order to capture some of the dynamical effects due to the tidal field, we also included a calculation of the shift in frequency, radius, energy and angular momentum  of the ISCO. 
In contrast with the conservative piece of the gravitational self-force, which always causes a positive frequency shift in Schwarzschild~\cite{BarackSago09}
and for all spins sampled in  Kerr~\cite{isoyama2014gravitational}, the tidal field could lead to either a positive or a negative ISCO frequency shift, depending on the inclination angle of the orbit. In particular, orbits with $\hat{n} \cdot \hat {\bf L} < 1/\sqrt{3}$ undergo a negative frequency shift, while orbits with $\hat{n} \cdot \hat {\bf L}> 1/\sqrt{3}$ undergo a positive frequency shift. 
%For this system
A negative frequency shift corresponds to an earlier  merger, and a positive frequency shift to a later merger. 

To the best of our knowledge, our analysis is the first fully-relativistic one of three-body systems.
In the future, it will be interesting to extend our analysis to the cases of a central Kerr black hole and of inclusion of  the gravitational self-force.
In addition, for planetary systems it has been shown that octupole-order tidal field by the third body could generate much richer dynamics \cite{li2015implications,katz2011long}.
For the system we consider here, tidal effect affects the GW waveform mostly through transient resonance phases. During such limited evolution within transient resonances, as the magnitude of octupole order tidal force is $\mathcal{O}\left ( \frac{r_0}{d}\right )^2$ smaller than the quadrupole order tidal force, it should be subdominant unless we are dealing with highly-eccentric orbits.

Finally, we note that we have analyzed the case where the inner binary is in the extreme mass-ratio regime. 
It is reasonable to expect that  the analytical understanding that we have provided  could shed some light
 on the dynamics of triple systems with a comparable-mass (stellar mass) inner binary,
  similarly to the spirit of using the Effective-One-Body formalism for describing the nonlinear two-body problem~\cite{PhysRevD.59.084006}. 
  Such triple systems could form in nuclear field clusters and they
  are expected to be important sources for {\it ground-}based GW detectors~\cite{antonini2012secular,thompson2011accelerating,antonini2014black,silsbee2016lidov,0004-637X-834-2-200,wen2003eccentricity}.

%We note that it is conceptually trivial (although by no mens technically trivial) to add self-force effects to our three-body system.

%---------------------------------------------------------------------------------------------------------------------------------------------------------------------------------------------------------------------
%---------------------------------------------------------------------------------------------------------------------------------------------------------------------------------------------------------------------

{\it Acknowledgements-} H.Y. thanks Scott Hughes for valuable discussions and comments, as well as Scott Tremaine and Chiara Mingarelli for information on evolution history of MBH binaries. M.C. acknowledges partial financial support by CNPq (Brazil), process number 308556/2014-3. 
The authors thank anonymous referees for interesting discussions and many helpful comments.

%---------------------------------------------------------------------------------------------------------------------------------------------------------------------------------------------------------------------
%---------------------------------------------------------------------------------------------------------------------------------------------------------------------------------------------------------------------

\bibliography{References}

\end{document}